# The use of financial and sustainability ratios to map a sector. An approach using compositional data

Elena Rondós-Casas[1], Germà Coenders[1,*], Miquel Carreras-Simó[1], Núria Arimany-Serrat[2]

**Purpose:** The article aims to visualise in a single graph fish and meat processing company groups in Spain with respect to long-term solvency, energy, waste and water intensity and gender employment gap.

**Design/methodology/approach:** The selected financial, environmental and social indicators are ratios, which require specific statistical analysis methods to prevent severe skewness and outliers. We use the compositional data methodology and the principal-component analysis biplot.

**Findings:** Fish-processing companies have more homogeneous financial, environmental and social performance than their meat-processing counterparts. Specific company groups in both sectors can be identified as poor performers in some of the indicators. Firms with higher solvency tend to be less efficient in energy and water use. Two clusters of company groups with similar performances are identified.

**Research limitations/implications:** As of now, few firms publish reports according to the EU Corporate Sustainability Reporting Directive. In future research larger samples will be available.

**Social Implications:** Firm groups can visually see their areas of improvement in their financial, environmental and social performance compared to their competitors in the sector.

**Originality/value:** This is the first time in which visualization tools have combined financial, environmental and social indicators. All individual firms can be visually ordered along all indicators simultaneously.

---

[1] University of Girona, Spain.

[2] University of Vic-Central University of Catalonia, Spain

[*] Correspondence: germa.coenders@udg.edu

# 1 Introduction

The Sustainable Development Goals (SDG) of the 2030 Agenda and the European Green Deal seek sustainable development and climate neutrality by 2050, and to decarbonize several sectors of activity. The agri-food sector and companies that process, preserve and produce meat, as well as companies that process and preserve fish, crustaceans and mollusks, are no exception and are experiencing changes at a financial and sustainability level. In addition, the linear business models of these companies are migrating towards circular models, according to Environmental, Social and Governance (ESG) criteria, to reduce the carbon footprint and lower Greenhouse Gas (GHG) emissions with environmental actions focused on more efficient management of water, energy and waste (Cantino et al., 2017; Lindkvist & Saric, 2020; Yilmaz, 2022), and social actions in this regard (Siew, 2023).

This study assesses the financial solvency of the companies in fish and meat processing, as well as the consumption of energy, water and waste, for each million euros of net income at an environmental level, together with an analysis of employment by gender. The aim is to graphically visualize the performance of these companies at a solvency level, at an environmental level and at a gender level. The study shows the annual financial, environmental and social health of the sector to diagnose how integrated information, according to European regulations, affects decision-making in the sector. The regulations used are the European Sustainability Reporting Standards (ESRS), in accordance with the EU Corporate Sustainability Reporting Directive (CSRD) 2022/2464 and the Global Reporting Initiative (GRI), which allows for better harmonization of sustainability information with respect to the previous Directive 2014/95/EU of the European Parliament and of the Council (NFRD).

It should be noted that food processing companies have a significant environmental impact that manifests itself in various areas, such as water pollution, GHG emissions, deforestation and the generation of solid waste and must prepare mandatory or voluntary sustainability reports (Anguiano-Santos & Rodríguez-Entrena, 2024; Villacampa-Porta et al., 2025). Food sector corporations have been linked to the discharge of toxic substances into aquatic ecosystems, the loss of biodiversity due to intensive production practices and the massive use of natural resources. On the other hand, food waste represents a significant energy and water expenditure for products that are not consumed (IRTA - Institut de Recerca i Tecnologia Agroalimentàries, 2024). These dynamics highlight the need to implement stricter environmental policies and promote more sustainable production and consumption models within the food sector (Forética, 2025; Villacampa-Porta et al., 2025), integrating financial and sustainability information (Anguiano-Santos & Rodríguez-Entrena, 2024). Although, in recent years, companies in the agri-food sector have increased the integration of financial information with ESG indicators, studies such as that of Anguiano-Santos & Rodríguez-Entrena in 2024, indicate little ESG disclosure according to the regulations of the NFRD Directive. The new CSRD Directive has facilitated the harmonization of integrated information according to ESRS standards, and studies are needed to assess the transparency and compliance with this regulation. Specifically, a European study shows that companies with digital maturity, good financial management and ESG commitment have better overall performance thanks to integrated information (Marczewska et al., 2025).

The theoretical framework highlights that food processing companies adopt comprehensive strategies to reduce environmental impact, using technological innovations, efficient resource management and circular economy models, to promote the use of renewable energy, lower water consumption and reduce waste (FAO, 2019; Gustavsson et al., 2011). Also, circularity and digitalization strategies show how traceability and waste management are integrated into the value chain (Geissdoerfer et al., 2017). In parallel, ecodesign and the quantification of environmental footprints are promoted to improve the efficiency of processes (Alvarez et al., 2014). However, the literature also points out the risk of greenwashing, highlighting the importance of independent assessment (Delmas & Burbano, 2011). Therefore, the sector shows a trend towards more sustainable production models, to reduce emissions, have efficient water consumption and carry out correct waste management, with gender equality in the value chain.

In this research, the objective is to relate the long-term solvency ratio to sustainability performance (Cantino et al., 2017; Lindkvist & Saric, 2020; Yilmaz, 2022), with some ratios defined according to Directive 2022/2464 (CSRD) and the 12 ESRS (Commission Delegated Regulation (EU) 2023/2772): *energy intensity* (total energy consumption per net revenue, megawatt-hours per million euros); *water intensity* (total water consumption in cubic meters per million euros of net revenue); *waste intensity* (company waste generation in tons per million euros) and *gender employment gap* (proportion of male and female workers). Specifically, the sustainability performance indicators considered in this study include relevant examples from the ESRS regulation focused on key environmental aspects, such as climate change, pollution, water resources management, biodiversity and the circular economy with a common denominator in relation to net income. The ESRS regulation in the social field highlights transparency in the proportion between men and women in the value chain.

It is worth noting that the integration of accounting information and ESG metrics has advanced with the establishment of the International Sustainability Standards Board (ISSB) under the auspices of the IFRS Foundation, which has published the IFRS S1 (general sustainability disclosures) and S2 (climate disclosures) standards with the aim of unifying and making sustainability disclosures comparable in accordance with traditional financial reports (IFRS Foundation, 2025a, 2025b). As of July 2025, more than 30 jurisdictions have these standards in the regulatory framework (KPMG, 2025), to ensure the reliability of ESG information in accounting processes and strategic decision-making (Zhou et al., 2025).

As a joint visualization tool for the financial and ESG ratios we use the principal component analysis biplot. To solve the problems of asymmetry and outliers of ratios in statistical analysis (Frecka & Hopwood, 1983; Iotti, Ferri, et al., 2024; Iotti, Manghi, et al., 2024; Isles, 2020; Linares-Mustarós et al., 2018, 2022) the biplot is drawn with the Compositional Data (CoDa) methodology as has been done for financial ratios (Carreras-Simó & Coenders, 2020; Saus Sala et al., 2023; Saus–Sala et al., 2021) and ESG Ratios (Todorov & Simonacci, 2020) separately.

This research is in line with the SDG of the United Nations of the 2030 Agenda, and in line with the European Green Deal (2019) to achieve climate neutrality by 2050. Specifically, it reflects how polluting industries must migrate to circular models for the decarbonization of the agri-food sector according to the environmental standards defined in the ESRS of the EU CSRD 2022/2464.

After this introduction, the methodological part and the results reached allow us to present a reasoned discussion to conclude what the relationship between solvency and sustainable performance is in the sector under study, as a basis for integrated information to make the appropriate economic decisions in line with the regulations of the field.

## 2 Methodology

### 2.1 Use and limitations of financial ratios in business analysis

Since the late 1800s, business analysis has employed ratios (Horrigan, 1968), a practice with both supporters and critics. The idea that a simple division provides valuable information for analysis or prediction (Barnes, 1987) has led to their use for many purposes.

In general, financial analysis using ratios can be divided into two main groups of studies: those that assess the economic and financial situation of a company, and prediction models. In the assessment domain, ratios are used as a tool for decision making. Examples can be found in the analysis of liquidity (Herrera Freire et al., 2017), analysis of corporate profitability (Qin et al., 2022; Trejo-Pech et al., 2023), debt capital (Iotti, Manghi, et al., 2024), equity valuation (Nissim & Penman, 2001) or a combination of all of them (Piotroski, 2000). Examples of prediction models include those dealing with bankruptcy or failure (Altman, 1968; Beaver, 1966; Dimitras et al., 1999; Hillegeist et al., 2004; Soukal et al., 2024; Staňková & Hampel, 2023; Veganzones & Severin, 2021; Wilson & Sharda, 1994), sometimes combined with credit risk (do Prado et al., 2016). They are also used to predict fraud detection (Amat Salas, 2020) or stock market returns (Dimitras et al., 1999).

Ratios have been used to analyse one or more companies, or to directly examine a sector, sub-sector or companies in a particular geographical location. Examples include studies of small firms (Tascón et al., 2018), dotcom companies (Ravisankar et al., 2010), Belgian manufacturing industries (Buijink & Jegers, 1986) or biogas sector (Iotti, Manghi, et al., 2024).

In the current business environment, financial ratios alone are insufficient for a comprehensive company analysis. It is also necessary to consider other key performance indicators, including efficiency, sustainability, impact, engagement and innovation (Siew, 2023). These new indicators, which are also ratios, allow the degree of compliance to be monitored and comparisons to be made between companies. Ratios are sufficiently objective to satisfy all interest groups (Denčić-Mihajlov & Zeranski, 2017).

Ratios are a proven tool for the analysis of one company, for the comparisons between very few companies. When applied to a data set, the use of statistical tools leads to a few limitations.

- It is difficult to achieve symmetry with the raw values of financial ratios. Statistically, asymmetry leads to invalid inferences about the population being studied because many statistical methods require normal distributions to produce valid results (Faello, 2015; Frecka & Hopwood, 1983; Iotti, Manghi, et al., 2024; Linares-Mustarós et al., 2018; Oktaviano et al., 2024; Trejo-Pech et al., 2023).

- Outliers are inconsistent observations that have a large effect on the statistical results. The reason for this can not only be that the data comes from an atypical company, but also that the ratio denominator is close to zero, resulting in an unreasonably high ratio. (Faello, 2015; Frecka & Hopwood, 1983; Linares-Mustarós et al., 2018).

- Finally, results are inconsistent when permuting the denominator and the numerator of the ratio, for instance when considering assets over liabilities -solvency- instead of liabilities over assets -indebtedness (Coenders, Sgorla, et al., 2023; Frecka & Hopwood, 1983; Linares-Mustarós et al., 2022)

### 2.2 Overcoming limitations and the application of compositional data

The prevailing approach to the ratio problems is to identify solutions for each individual case. For outliers, the challenge lies in both their detection (Frecka & Hopwood, 1983; Sullivan et al., 2021) and treatment. Simple methods that have been commonly used include their elimination or winsorizing (Deshpande, 2023; Lev & Sunder, 1979). More complex methods such as Mahalanobis distances (Sullivan et al., 2021) have also been employed.

In terms of techniques to achieve symmetry, there are several options to consider. These include data transformation techniques, such as square roots or cubic roots (Deakin, 1976; Martikainen et al., 1995), power transformations (Box-Cox) (Watson, 1990) and rank transformations (Kane et al., 1998), among others.

A variety of methodologies have been employed in the treatment of violations of normality. These include non-parametric methods (Iotti, Ferri, et al., 2024; Iotti, Manghi, et al., 2024), and robust standard errors (van der Heijden, 2011).

Despite the existence of partial solutions to the ratio issues, it was of significant interest to researchers and professionals to identify a technique that could resolve all the issues simultaneously. This objective was achieved through the implementation of a well-developed toolbox known as Compositional Data (CoDa) Analysis (Aitchison, 1982; Coenders, Egozcue, et al., 2023), which originated in the STEM scientific fields but has since been applied to numerous other areas, including financial analysis (Arimany Serrat et al., 2022; Coenders, Sgorla, et al., 2023; Fry, 2011; Magrini, 2025)

The employment of the CoDa methodology has been demonstrated to facilitate the elimination of data asymmetry and outliers and to enable the permutation of the numerator and denominator of the ratios. (Carreras-Simó & Coenders, 2020). Furthermore, the methodology is applicable to different statistical methods, parametric and non-parametric, with and without robust standard errors, and with a full non-trimmed sample, which represents a significant advantage.

The starting point of the CoDa methodology is examining a sample of strictly positive parts from a relative point of view.

$$x = (x_1, x_2, \ldots, x_D) \text{ with } x_j > 0, j = 1, 2, \ldots, D \qquad (1)$$

Its objective is thus the same as in financial ratio analysis.

## 2.3 Selection of variables

The next step is to select the variables to be analysed and adapt them to the CoDa methodology. The idea of our study is in line with the idea of Triple Bottom Line (TBL) reporting (Elkington, 1998; Nogueira et al., 2023; Zaharia & Zaharia, 2021), which recognises that to achieve corporate sustainability we need to link the economic, social and environmental roles of the company. The selection of indicators for each company will therefore be a mixture of financial and sustainability indicators. There are many potential such indicators and what follows is only a simple example aimed at the illustration of the proposed methodology.

In the example which follows and in terms of financial indicators, net revenue, total assets and total liabilities have been selected, while the rest are indicators that allow us to examine sustainability issues. In terms of sustainability, some indicators have been selected related to climate change, water and marine resources, circular economy and employee genders distribution in the value chain.

All indicators are represented by the $x_j$ variables, which are the $D=8$ positive financial and sustainability categories. These eight indicators are needed to calculate the financial and ESG ratios analysed in this example:

- $x_1$: net revenue,
- $x_2$: total assets,
- $x_3$: total liabilities,
- $x_4$: energy consumption,
- $x_5$: water consumption,
- $x_6$: waste generation,

- $x_7$: male employees,
- $x_8$: female employees.

The financial ratio we will use is the solvency ratio because it is related to capital structure, is critical in its relationship to sustainability performance, and can be affected by the investments needed to improve said performance (Cantino et al., 2017; Lindkvist & Saric, 2020; Yilmaz, 2022).

- The solvency ratio measures the extent of a company's leverage:

$$Solvency\ ratio\ =\ total\ assets/total\ liabilities\ =\ x_2/x_3 \qquad (2)$$

Sustainability has been measured in a variety of ways, including semi-structured interviews (Khan & Quaddus, 2015), the use of a balanced scorecard (Vu et al., 2017), an integrated model (Castro & Chousa, 2006) or a questionnaire survey (Andersson et al., 2022). In our work, we have chosen to analyse sustainability using ratios as key performance indicators (Siew, 2023).

Following the implementation of the CSRD (2022/2464) and the adoption of the ESRS (Commission Delegated Regulation (EU) 2023/2772), companies will transition to reporting sustainable indicators in a unified manner. The ESRS represent a practical implementation tool, organised in 12 standards. Each standard includes information on governance, strategy, impact, risk and opportunity management across the value chain, as well as metrics and guidance on disclosure targets for each topic. In addition, there is a section on calculation guidelines, which contains ratios.

- Energy intensity: Included in the ERSR under the climate change theme and the energy sub-theme, it assesses the efficiency of energy use. It is measured as total energy consumption per net income and the units are megawatt hours per million EUR.

$$Energy\ intensity\ =\ energy\ consumption/net\ revenue\ =\ x_4/x_1 \qquad (3)$$

- Water intensity: Classified in ERSR under the theme of water and marine resources and the sub-theme of water use, it measures efficiency in the use of clean water and sanitation. This key performance indicator is measured as total water consumption in cubic metres per million EUR in net revenue.

$$Water\ intensity\ =\ water\ consumption/net\ revenue\ =\ x_5/x_1 \qquad (4)$$

- Waste intensity: Classified in ERSR under the circular economy theme and the waste sub-theme, it is a measurement of the company's waste generation. The unit of measurement is tons per million EUR.

$$Waste\ intensity\ =\ waste\ generation/net\ revenue\ =\ x_6/x_1 \qquad (5)$$

- Gender employment gap: Classified in ERSR under the theme of workers in the value chain and the sub-theme of equal treatment and opportunities for all, it measures the ratio of male and female empoyees.

$$Gender\ employement\ gap\ =\ male\ employees/female\ employees\ =\ x_7/x_8 \qquad (6)$$

### 2.4 Log-ratio transformations

The use of compositional data does not remove the need for interpretation and visualisation, as would be the case with any other statistical data. However, it does mean that the input data must be prepared according to the CoDa methodology for visualization purposes (Saus Sala et al., 2023; Todorov & Simonacci, 2020). Computing logarithms of the ratios (log-ratios) is the first necessary transformation of the input data (Pawlowsky-Glahn et al., 2015). The simplest form of log-ratios is the so-called pairwise log-ratios, computed between two indicators, as in equations (2) to (6):

$$log\left(\frac{x_i}{x_j}\right) \quad (7)$$

After log-ratio transformation, the data is not affected by scale variations, has no asymmetry and is close to a normal distribution (Aitchison, 1986; Linares-Mustarós et al., 2018, 2022; Pawlowsky-Glahn et al., 2015). Besides, log-ratios are invariant to numerator and denominator permutation:

$$log\left(\frac{x_i}{x_j}\right) = \log(x_i) - \log(x_j) = -(\log(x_i) - \log(x_j)) = -log\left(\frac{x_j}{x_i}\right) \quad (8)$$

A particularly useful type of log-ratio employed in CoDa is the centred log-ratio or CLR (Aitchison, 1983). Although the CLR transformation has no direct interpretation, it is highly useful in multivariate analysis and can be adapted to visualisation tools such as the biplot, which will be discussed in more detail later. The process involves the obtaining of a centred log-ratio for each variable studied, which in our case will be eight. For each of these, the CLR will relate the variable to the geometric mean of the total set.

$$CLR_1 = log\left(\frac{x_1}{\sqrt[8]{x_1 x_2 \ldots x_8}}\right)$$

$$CLR_2 = log\left(\frac{x_2}{\sqrt[8]{x_1 x_2 \ldots x_8}}\right) \quad (9)$$

$$\ldots$$

$$CLR_8 = log\left(\frac{x_8}{\sqrt[8]{x_1 x_2 \ldots x_8}}\right)$$

## 2.5 Data collection

The data refers to 2023 and was obtained using three tools: the Sabi INFORMA (Iberian Balance sheet Analysis System, accessible at https://sabi.bvdinfo.com) database, the companies' websites and the annual accounts and management reports. The financial indicators required ($x_1$: net revenue, $x_2$: total assets and $x_3$: total liabilities) can be obtained from the annual accounts that companies are obliged to file annually with the Commercial Registry and which are public.

The remaining indicators ($x_4$: energy consumption; $x_5$: water consumption; $x_6$: waste generation; $x_7$: male employees; $x_8$: female employees) are non-financial variables that can be obtained from companies' environmental, social and governance (ESG) reports. The ESG reports should include information on environmental, social and personnel issues, human rights, anti-corruption and anti-bribery, and commitment to transparency, according to European Commission guidelines and GRI standards. However, the format, detail and depth of data vary significantly between companies. This situation will be resolved once Directive (EU) 2022/2464 is transposed and companies submit a sustainability report in accordance with the ESRS.

Until 2017, Spanish companies had the option of disclosing their non-financial data. However, following the transposition of Directive 2014/95/EU in 2018, this became mandatory for certain companies. According to Spanish Law 11/2018, companies are required to file the ESG reports if falling under the following criteria:

a) The average number of employees is more than 250.
b) They are public interest entities, or that for two consecutive financial years they meet at least two of the following circumstances at the closing date:
    a. The total assets of the company exceed EUR 20 MM.
    b. The net revenue exceeds EUR 40 MM.
    c. The average number of workers employed during the financial year exceeds 250.

In the case of companies that form part of a corporate group, the requirements are based on the same criteria and amounts but referring to the employees of the group and to the total consolidated assets and net revenue. Individual companies have the option of either including a statement of non-financial information in the management report or producing a separate report. Groups, on the other hand, are required to provide a consolidated annual report that includes both financial and non-financial information. For our reporting needs, the consolidated annual report filed with the Commercial Register was the most appropriate document, as it has the advantage that many groups present all information we need, financial and non-financial. The disadvantage is that we must manually extract the relevant information.

### 2.5.1 Sabi INFORMA

The Sabi INFORMA tool is a software solution that facilitates the analysis of financial data from Spanish and Portuguese companies. Its advanced search system, complete with filters, allows users to specify the necessary restrictions to obtain data tailored to their study's requirements. It also allows users to consult the annual accounts filed by companies in the Commercial Registry.

The following restrictions were applied:

- The companies were active,
- operated within the following sectors, as classified according to the NACE codes 101X (Processing and preserving of meat and production of meat products) and 102X (Processing and preserving of fish, crustaceans and molluscs),
- submitted consolidated annual accounts,
- and fulfilled the requirements for submitting the consolidated management report.

Following a thorough evaluation of the established requirements, 20 company groups belonging to the meat production sector and 8 to the fish and seafood sector were identified. Once the resulting list of companies has been obtained, it is possible to define a report with the necessary variables. However, in our case, due to the requirement for non-financial data, an individual data search was necessary.

### 2.5.2 Annual accounts and management reports filed with the Commercial Registry.

Non-financial information is subject to the same regime of formulation, approval and publication in the Commercial Registry as the annual accounts. For this reason, we were able to consult the deposited consolidated management reports of the above company groups to obtain the eight financial and non-financial indicators we needed. The parent company is obliged to prepare the consolidated non-financial information statement, including all its subsidiaries and for all the countries in which it operates.

Unfortunately, we were unable to obtain non-financial data for all the selected companies. This was because some companies did not have updated data, and others expressed certain non-financial indicators with percentage increases or decreases, rather than actual data for the year. The final sample sizes were 11 company groups in the meat sector (44.8 % of the sector's total net revenue) and 6 in the fish sector (59.5 % of the sector's total net revenue). There was also a need for harmonization of the data, as some companies had different measurement units for their sustainability indicators.

We also conducted a thorough review of the websites of all relevant company groups but were unable to identify additional data beyond what was previously outlined. The descriptive statistics are in Table 1.

|  | Mean | SD | Minimum | Q1 | Median | Q3 | Maximum |
|---|---|---|---|---|---|---|---|
| Net revenue (MM EUR) | 974 | 1.035 | 54 | 211 | 713 | 1.040 | 4.148 |
| Total assets (MM EUR) | 635 | 660 | 44 | 123 | 376 | 992 | 2.622 |
| Total liabilities (MM EUR) | 378 | 380 | 23 | 80 | 255 | 603 | 1.424 |
| Energy consumption (megawatt hours) | 195.489 | 238.643 | 6.269 | 42.281 | 111.003 | 212.310 | 878.780 |
| Water consumption ($m^3$) | 1.421.715 | 1.432.267 | 35.730 | 323.252 | 944.041 | 1.548.640 | 4.448.501 |
| Waste generation (metric tons) | 45.369 | 104.225 | 329 | 3.441 | 13.166 | 27.783 | 453.002 |
| Male employees | 1.371 | 1.278 | 82 | 408 | 608 | 1.949 | 4.964 |
| Female employees | 830 | 597 | 117 | 236 | 763 | 1.118 | 2.017 |

Table 1. Descriptive statistics for the *n*=17 companies in the sample

### 2.6 Compositional Biplot

Following the acquisition of the data, it was transformed into CLR to facilitate analysis from a compositional perspective. To visualize and interpret the results, a CoDa biplot was drawn from a principal component analysis (Aitchison, 1983; Greenacre et al., 2022).

This tool has been widely applied in the analysis of financial statements (Carreras-Simó & Coenders, 2020; Coenders, 2025; Coenders & Arimany-Serrat, 2025; Saus Sala et al., 2023; Saus–Sala et al., 2021). The biplot's key strength is its ability to simultaneously visualise all companies in the sector and all possible pairwise log-ratios. In this article we apply it for the first time to financial and non-financial company data, thus broadening the spectrum of analysis and applicability of the tool. The biplot is generated using CaDaPack2.03.06, which is available free of charge at https://ima.udg.edu/codapack/.

The companies under study in the CoDa biplot are represented by a point. The CLR corresponding to the eight indicators are plotted using rays. At the extreme end of these rays lies the financial or non-financial value in the numerator of the CLR.

An advantage of using the biplot in interpreting financial and ESG ratios is that it allows lines to be drawn between two extremes of any two indicators in order to obtain the pairwise ratio between the two. These additional lines, called links, make it possible to rank all the companies shown according to a any particular ratio like those in equations (2) to (6). To do this, the companies are projected orthogonally (i.e. at 90 degrees to the link) onto the line. Once projected onto the corresponding ratio, they are sorted according to their proximity to the numerator or denominator of the ratio displayed.

### 3 Results

The CoDa biplot in Figure 1 is the result of the study. The visualisation of the companies has been set so that companies in the meat production industry (NACE 101X) are coloured red, while those in the fish industry (NACE 102X) are coloured blue.

An initial analysis of the distribution of the coloured points is warranted. This analysis reveals that the blue-coloured company groups exhibit higher levels of concentration, while the red-coloured company groups demonstrate greater dispersion, indicating a lower degree of similarity between them.

By joining the extreme of the assets CLR ray with that of the liabilities CLR ray, we obtain the link representing the solvency ratio (assets/liabilities) in Equation 2, represented as a green line. The company groups are then orthogonally projected on this line. In this way, it is interpreted that those companies whose projection is closer to the asset side have a significantly high solvency ratio. Conversely, those closer to the liability side have a high indebtedness. Distance to the link does not matter. Company groups 11 and 17 stand out for the highest solvency and company group 6 for the lowest.

By joining the extreme of the energy and revenue rays we obtain the link representing the energy intensity ratio (energy/revenue) in Equation 3. The company groups projecting closer to the energy extreme have a significantly high energy intensity ratio, and those closer to the revenue ray have a lower energy intensity ratio. Company groups 11 and 17 stand out for the highest energy intensities and company group 6 for the lowest.

The link between the extremes of the water and revenue rays in Equation 4 yields the water intensity ratio (water consumption / revenue). The company groups projecting closer to the water extreme exhibit a higher water intensity ratio, whilst those approaching the revenue ray demonstrate a lower water intensity ratio. It is notable that companies in groups 17, 8 and 11 display the highest levels of water intensity, while those in groups 7 and 16 exhibit the lowest.

The link between the extremes of the waste and revenue rays represents the waste intensity ratio (waste/revenue) in equation 5. The company groups that are closer to the waste extreme have a significantly high waste intensity ratio, while those that are closer to the revenue ray have a lower waste intensity ratio. Company groups 8, 17 and 4 are distinguished by the highest levels of waste intensity, while groups 16 and 7 are notable for their comparatively low levels of waste intensity.

The gender employment gap (male employees/female employees) in equation 6 is represented by the link between the extremes of the male and female employee rays. The company groups with a greater propensity to have more male employees have a much more pronounced gender gap. Conversely, the company groups that are closer to the female extreme have a smaller gender gap, characterized by a higher proportion of female employees relative to male employees. It is noteworthy that company groups 17, 8 and 4 have the most pronounced gender gaps, while company groups 7 and 16 have the least pronounced gaps.

The biplot can be used to visually identify clusters of similarly behaved company groups. The company groups exhibiting minimal separation in the biplot are those that possess a similar financial and sustainability structure and, therefore, would form a similar operating cluster. The biplot reveals the presence of two distinct clusters. The first cluster (hereafter, cluster 1) comprises cases 2, 5, and 13, while the second cluster (hereafter, cluster 2) consists of points 14, 15, and 16. A comparative analysis reveals that cluster 1 includes two groups of companies from the 101X sector and one from the 102X sector, while in cluster 2 all three groups belong to the 102X sector.

Cluster 1 is very close to the origin of the rays, which indicates that its members have financial and ESG performance close to the sample average. Conversely, cluster 2 exhibits weaker solvency, lower intensity in terms of water and waste, and a reduced gender employment gap. In terms of energy intensity, cluster 2 behaves closely to the sample average.

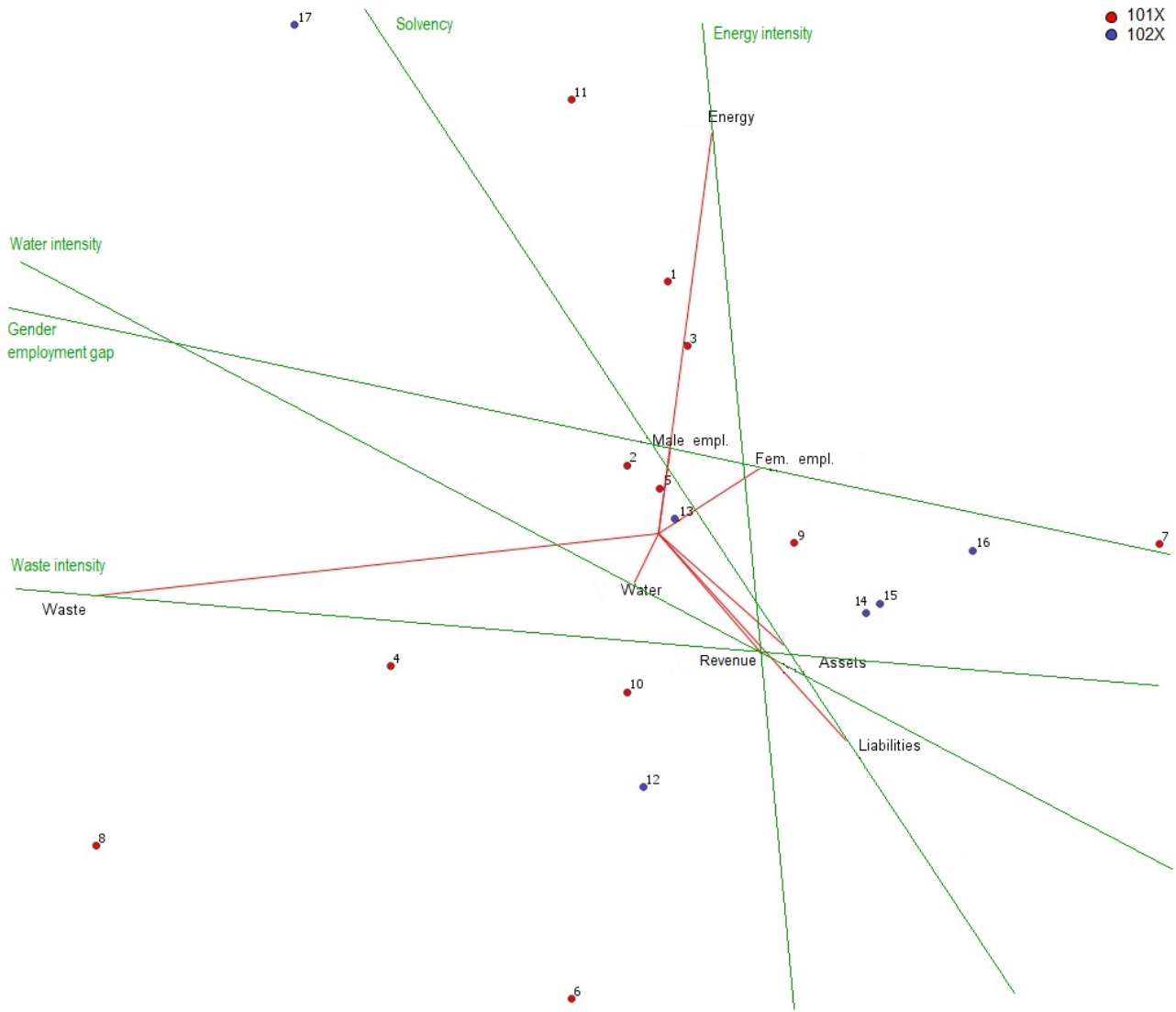

Figure 1. CoDa biplot of the financial and ERSR ratios on our example

## 4 Conclusions and discussion

The scientific and political communities agree on reducing $CO_2$ emissions to limit the magnitude and scope of climate change (Todorov & Simonacci, 2020). This research identifies energy, water and waste efficiency as a strategic need for the competitiveness and sustainability of the meat and fisheries sector, due to its resource intensity and ecological impact. Specifically, in terms of energy, the adoption of efficient refrigeration technologies, heat recovery and the integration of renewable energies reduce emissions and costs (Schröter, 2025). In terms of water, the implementation of recirculation systems and the reuse of wastewater with appropriate technologies help reduce the water footprint, especially in sectors such as meat, which is responsible for 75% of the water loss associated with food waste (Skawińska & Zalewski, 2022). Regarding the fish sector, there are studies that show an increase in the water footprint per ton produced (Pahlow et al., 2015), with the need to optimize refrigeration and washing systems. Regarding waste management, the valorization of by-products and the reduction of plastic packaging help to achieve circular economy models (Singh et al., 2022).

It is also relevant to consider the gender perspective in the meat and fish sectors to improve sustainability, productivity and innovation, as gender equality is directly linked to more inclusive, resilient and competitive societies on a global scale.

Once the results of the study have been presented and according to the analysis developed in the meat industry (in red) and the fish industry (in blue), considering the long-term solvency ratio (which measures a company's capacity to meet its long-term financial obligations); energy, water and waste intensity; and, the gender employment gap, using a CoDa biplot, the study shows that fish companies present higher levels of concentration, while meat companies show greater dispersion. This situation indicates a lower degree of similarity between these companies, despite belonging to the agri-food industry.

On the other hand, the study allows us to conclude that companies with greater long-term solvency, present a higher energy intensity and water intensity, especially in the case of the meat companies analyzed, with gender gaps in both types of industry.

Specifically, company groups 11 and 17, meat and canned fish respectively, stand out for their high solvency, and high energy and water intensities, in contrast to meat company group 6. It should be noted that company group 17 has ISO 14001 and ISO 45001 certifications and has received subsidies to improve energy efficiency with co-financing from the European Regional Development Fund (ERDF). Specifically, it received a subsidy co-financed by the European Agricultural Fund for Rural Development according to the Navarra Investment Plan 2014-2020. In addition, company group 17 has a high intensity in waste and a high gender gap. In spite of their certifications and subsidies, these two companies stand out for their need to continue improving their environmental efficiency.

Company group 6, which has a low solvency, has implemented renewable energy and thermal recovery systems, has the best energy intensity and needs more actions only with respect to waste intensity and the gender gap, whose investments should be financed by shareholders' equity in order to improve solvency.

Regarding water consumption intensity, meat company group 8 also presents high levels of consumption, unlike company groups 7 and 16, meat and fish respectively, which have lower levels. Regarding waste intensity, company groups 4, 8 and 17 are distinguished by high values, compared to company groups 7 and 16, which are lower. It is noteworthy that company groups 4, 8 and 17 present pronounced gender gaps, while company groups 7 and 16 do not present so much inequality. It is noteworthy that meat company group 8, despite having high water consumption, high waste generation and a high gender gap, has a commitment to sustainability and has installed self-consumption solar panels in all its production centers, generating more than 15 GWh per year and significantly reducing $CO_2$ emissions. At the same time, it has implemented measures to improve the efficiency of water use and has incorporated circular economy practices into its processes. It also promotes gender equality and regional development through a foundation, which supports vulnerable groups.

Meat company group 7 applies measures to optimize energy and water consumption in its production processes, as well as protocols for waste management and promotes gender equality initiatives. In Figure 1, water and waste intensity is low, as is the gender gap.

Therefore, the companies in the sample analyzed that strongly deviate from the average are mostly meat company groups (6, 7, 8, 11) and one canned fish company group (17). The study shows that Spanish food processing companies have different strategies in terms of sustainability. Specifically, company group 6 is committed to renewable energy and thermal recovery, as well as efficient water and waste management with a circular economy model recognized with external certifications. Fish company group 17 integrates sustainability into its logistics and production, with special emphasis on reducing its carbon footprint through rail transport and plans more gender equality policies. Company group 8 has promoted an ambitious solar self-consumption plan and has incorporated measures for water efficiency and circularity, while strengthening its social responsibility through its foundation. Finally, company group 7 communicates good practices in the meat sector, optimizing resources, waste management and initiatives for labor equality. The study highlights the high solvency of meat company group 11 and fish processing company group 17, unlike meat company group 6 which has low solvency.

The study also identifies two clusters, cluster 1, which integrates company groups 2, 5 and 13, with an average close to the sample in terms of solvency, energy intensity, water intensity, waste intensity and gender gap; and cluster 2 which integrates company groups 14, 15 and 16, with a weak solvency and a low intensity in water consumption and waste generation, along with a reduced gender gap.

In cluster 1, with respect to long-term solvency, all three member company groups have the operational capacity to finance the necessary environmental transitions. Regarding sustainability performance, company group 13 (canned fish) has more than 95% of waste destined for recycling and/or recovery. Company group 5 (meat) has a plan that sets measurable objectives (-10% water and emission reductions) and communicates progress regarding their own renewable energy and the reduction of GHG emissions from 2020. Finally, meat company group 2 communicates photovoltaic investments and decarbonization. Regarding long-term solvency, all three have the operational capacity to finance environmental transitions.

Regarding cluster 2, the three fish processing company groups that make it up show clear commitments to efficiency and social responsibility. Company group 14 addresses social management with equality programs without clear evidence of a gender gap. Company group 15 stands out for its track record in eco-efficiency and circularity (recycling plastic waste and reducing the carbon footprint). Finally, company group 16 focuses on sustainable fishing, and strengthens its resilience through certifications and more innovation in the value chain.

Methodologically, this article presents the first joint visualization of ESG and financial indicators. Since ESG indicators are also ratios (Todorov & Simonacci, 2020) the same CoDa biplot used for financial ratios (Carreras-Simó & Coenders, 2020; Saus Sala et al., 2023; Saus–Sala et al., 2021) can accommodate both. In just two dimensions company or company groups can be ordered according to the ratio between any two magnitudes, be they financial or ESG.

The research has limitations. The first two, small sample, and single-year data are dictated by the slow and progressive adoption of the ESRS in the sustainability reports. The third is the few financial and ESG indicators included in this first example. To this end, future lines of study will focus on sectors of activity that have larger samples with complete ESG information, data from different years, and more indicators, such as short-term solvency in terms of cash flow, profitability, GHG emissions, percentage of employees with disabilities, gender pay gap, and board's gender diversity, to assess this integrated information that is crucial for making business decisions.

**Declaration of Conflicting Interests**

The authors declare no potential conflicts of interest with respect to the research, authorship, and/or publication of this article.

**Funding**

This research was supported by Spanish Ministry of Science, Innovation and Universities MCIN/AEI/10.13039/501100011033 and ERDF-a way of making Europe [grant number PID2021-123833OB-I00]; the Spanish Ministry of Health [grant number CIBERCB06/02/1002]; the Department of Research and Universities of Generalitat de Catalunya [grant numbers 2021SGR01197 and 2021SGR00403]; and the Department of Research and Universities, AGAUR and the Department of Climate Action, Food and Rural Agenda of Generalitat de Catalunya [grant number 2023-CLIMA-00037].